\documentclass[12pt]{article}
\usepackage{epsfig}

\begin{document}
\baselineskip=18pt

\begin{titlepage}
\begin{flushright}
OSU--HEP--08-05\\
UMD-PP-08-011
\end{flushright}
\vskip 2cm
\begin{center}
{\large\bf Minimal Supersymmetric Left--Right Model \\[0.03in]
with Automatic $R$--Parity}
\vskip 1cm
{\normalsize\bf
K.S.\ Babu$\,{}^1$ and Rabindra N. Mohapatra$\,{}^{2}$} \\
\vskip 0.5cm
{\it ${}^1\,$Department of Physics, Oklahoma State University,\\
Stillwater, OK~~74078, USA\\ [0.1truecm]
${}^2\,$Maryland Center for Fundamental Physics and Department of
Physics,\\ University of Maryland, College Park, MD 20742, USA \\[0.1truecm]
}

\end{center}
\vskip .5cm

\begin{abstract}

We revisit the minimal supersymmetric left--right model with
$B-L=2$ triplet Higgs fields and show that a self--consistent
picture emerges with automatic $R$--parity conservation even in
the absence of higher dimensional operators. By computing the
effective potential for the Higgs system including heavy Majorana
neutrino Yukawa couplings we show that the global minimum of the
model can lie in the charge and $R$--parity conserving domain. The
model provides natural solutions to the SUSY phase problem and the
strong CP problem and makes several interesting predictions. Quark
mixing angles arise only after radiative corrections from the
lepton sector are taken into account. A pair of doubly charged
Higgs fields remain light below TeV with one field acquiring its
mass entirely via renormalization group corrections. We find this
mass to be not much above the Bino mass. In the supergravity
framework for SUSY breaking, we also find similar upper limits on
the stau masses. Natural solutions to the $\mu$ problem and the
SUSY CP problem entails light $SU(2)_L$ triplet Higgs fields,
leading to rich  collider phenomenology.

\end{abstract}

\end{titlepage}

\setcounter{footnote}{0}
\setcounter{page}{1}
\setcounter{section}{0}
\setcounter{subsection}{0}
\setcounter{subsubsection}{0}

\def\spur#1{\mathord{\not\mathrel{{\mathrel{#1}}}}}
\def\ol#1{\overline{#1}}


\section{Introduction}

Left-right symmetric extensions of the Standard Model (SM) based
on the gauge group \cite{patimohap} $SU(3)_C\times SU(2)_L\times
SU(2)_R\times U(1)_{B-L}$ have many attractive features.  These
include an understanding of the origin of parity violation, and a
compelling rationale for small neutrino masses via the seesaw
mechanism.  The enlarged gauge symmetry allows for parity to be
defined as an exact symmetry, which is broken only spontaneously.
Right--handed neutrino is required to exist in order to complete
the $SU(2)_R$ multiplet, and so neutrino mass is natural.  In the
domain of flavor physics, the supersymmetric version of this
theory  resolves several problems of the popular minimal
supersymmetric standard model (MSSM): {\bf (i)} $R$--parity
emerges as an exact symmetry of MSSM, preventing rapid proton
decay and providing a naturally stable dark matter candidate
\cite{rnm}. This is possible if the $SU(2)_R\times U(1)_{B-L}$
gauge symmetry is broken down to $U(1)_Y$ by Higgs triplet fields
carrying  $B-L=\pm 2$. $R$--parity, which is part of the original
$B-L$ symmetry, will remain unbroken even after symmetry breaking
in this case. {\bf (ii)} It solves the SUSY CP problem
\cite{rasin,babu} because of parity invariance. Parity makes the
Yukawa couplings and the corresponding SUSY breaking $A$ terms
hermitian, and the gluino mass and the $\mu$ term real.  The
electric dipole moments of fermions will then vanish at the scale
of parity restoration. {\bf (iii)} Finally, it has all the
ingredients necessary to solve the strong CP problem without the
need for an axion, again by virtue of parity symmetry
\cite{rasin,babu}.  This is achieved by ensuring that the quark
mass matrix has a real determinant, which is possible since the
Yukawa couplings are hermitian.

Previous studies of this model focussed on two versions: (i) A TeV
scale version where $R$--parity is spontaneously broken by the
vacuum expectation value (VEV) of the right--handed sneutrino
\cite{kuchi}, or alternatively (ii) an $R$--parity conserving
version \cite{goran,chacko} where non-renormalizable (NR) higher
dimensional operators were included and played an essential role.
The reason for considering only these two versions is that in the
absence of the above features, i.e., $\left\langle \tilde{\nu}^c
\right\rangle \neq 0$, or the presence of NR operators, the global
minimum of the theory that is both $R$--parity conserving and
parity violating, breaks electric charge and is therefore
unacceptable. In the first version with $\left\langle
\tilde{\nu}^c \right\rangle \neq 0$, the $W_R$ scale must
necessarily be in the TeV range \cite{kuchi}, whereas in the
second one, it is necessarily above $10^{11}$ GeV. In the first
version, SUSY dark matter candidate is lost.  In the second
version, the possibility of solving strong CP problem via parity
symmetry is eliminated due to the essential presence of higher
dimensional operators which makes $\overline{\theta}$ large. It is
also difficult to solve the SUSY phase problem, since these higher
dimensional operators typically generate parity violating effects
in the fermion mass matrices.  Extensions of the minimal model which use
additional Higgs multiplets have been proposed.  Ref. \cite{bdm} introduces
Higgs doublets in addition to triplets, but
in such models $R$--parity conservation is exterior to parity
symmetry.  In Ref. \cite{benakli} $B-L=0$ Higgs triplets are introduced in addition
to the $B-L = \pm 2$ triplets, which is clearly non--minimal.

In this note we revisit the minimal SUSYLR model with $B-L=2$
Higgs triplets.  We assume that the higher dimensional operators
are absent or small, so that the solutions  to the strong CP and
the SUSY phase problems are still intact. The global minimum of
the tree--level Higgs potential is either charge violating, or
$R$--parity violating, as noted.  However, we find that inclusion
of the heavy Majorana neutrino Yukawa couplings in the effective
potential automatically cures this problem.  The vacuum that
preserves both electric charge and $R$--parity can naturally be
the global minimum of the full potential. We study the
consequences of such a setup.\footnote{It is not strictly required
that the vacuum we live in correspond to the global minimum of the
potential. Metastable vacua are acceptable, provided that the
tunnelling rate from that vacuum to the true vacuum is
sufficiently slow in comparison to the age of the Universe.}

The main results of our investigation can be summarized as
follows: {\bf (i)} In this general class of models, there are two
doubly charged Higgs and Higgsino fields with masses below a TeV.
One combination of these doubly charged Higgs boson fields has a
vanishing mass at the scale of $SU(2)_R \times U(1)_{B-L}$
breaking (denoted as $v_R$).  So its mass is calculable, arising
through renormalization group effects between $v_R$ and the weak
scale. We find its  squared mass to be positive with the Higgs
boson having a mass close to the Bino mass. {\bf (ii)} There exist
two pairs of Higgs doublets in the low energy, although one pair
is unlikely to be observed directly at the LHC. This naturally
leads to calculable flavor violation, which are within
experimental limits. {\bf (iii)} Renormalization group evolution
plays a crucial role in the generation of quark mixing angles.  In
fact, an asymmetry in the $\mu$ terms of the Higgs doublets
generated by the leptonic Yukawa couplings is what induces CKM
mixings. {\bf (iv)} In the version that solves the SUSY phase and
the strong CP problems and which provides an understanding of the
$\mu$ problem, there are also light $SU(2)_L$ triplet superfields
with TeV to sub-TeV scale masses with interesting collider
signature \cite{huitu,han}. These fields couple to left--handed
leptons with the couplings proportional to the heavy Majorana
neutrino masses.

\section{The basic structure of the model}

Quarks and leptons in the model have the following left-right
symmetric assignment under the $SU(3)_C \times SU(2)_L \times
SU(2)_R \times (1)_{B-L}$ gauge group.
\begin{eqnarray}
Q(3,2,1,{1 \over 3}) &=& \left(\matrix{u \cr d}\right);~~~~
Q^c(3^*,1,2,-{1 \over 3}) = \left(\matrix{d^c \cr -u^c}\right)
\nonumber \\
L(1,2,1,-1) &=& \left(\matrix{\nu_e \cr e}\right);~~~~
L^c(1,1,2,1) = \left(\matrix{e^c \cr -\nu_e^c}\right)~.
\end{eqnarray}
The minimal Higgs sector
consists of the following superfields:
\begin{eqnarray}
\Delta(1,3,1,2) &=&\left(\matrix{{\delta^{+} \over \sqrt{2}} & \delta^{++} \cr \delta^{0} & -{\delta^+ \over \sqrt{2}}}\right);~~
\overline{\Delta}(1,3,1,-2) = \left(\matrix{{\overline{\delta}^{-} \over \sqrt{2}} & \overline{\delta}^{0} \cr \overline{\delta}^{--} & -{\overline{\delta}^{-} \over \sqrt{2}}}\right);~~ \nonumber \\
\Delta^c(1,1,3,-2) &=& \left(\matrix{{\delta^{c^-} \over \sqrt{2}} & \delta^{c^0} \cr \delta^{c^{--}} & -{\delta^{c^-} \over \sqrt{2}}}\right);~~
\overline{\Delta^c}(1,1,3,2) = \left(\matrix{{\overline{\delta}^{c^+} \over \sqrt{2}} & \overline{\delta}^{c^{++}} \cr \overline{\delta}^{c^0} & -{\overline{\delta}^{c^+} \over \sqrt{2}}}\right);~~ \nonumber \\
\Phi_a(1,2,2,0) &=& \left(\matrix{\phi_1^+ & \phi_2^0 \cr \phi_1^0
& \phi_2^-}\right)_a~~(a=1-2);~~~S(1,1,1,0)~~.
\end{eqnarray}
This is the minimal Higgs system in the following sense.  The $(\Delta^c + \overline{\Delta^c})$ fields are needed
for $SU(2)_R \times U(1)_{B-L}$ symmetry breaking without inducing $R$--parity violating couplings.  The
$(\Delta + \overline{\Delta})$ fields are their left--handed partners needed for parity invariance.  Two
bidoublet fields $\Phi_a$ are needed in order to generate quark and lepton masses and CKM mixings.  The singlet
field $S$ is introduced so that $SU(2)_R \times U(1)_{B-L}$ symmetry breaking occurs in the supersymmetric limit.

The superpotential of the model is given by
\begin{eqnarray}
 W &=& Y_u Q^T \tau_2 \Phi_1 \tau_2 Q^c +  Y_d Q^T \tau_2 \Phi_2 \tau_2
Q^c + Y_\nu L^T \tau_2 \Phi_1 \tau_2 L^c  + Y_\ell L^T \tau_2
\Phi_2 \tau_2
L^c \nonumber \\
&+& i  \left(f^* L^T \tau_2 \Delta L + f L^{cT} \tau_2 \Delta^c
L^c
\right) \nonumber \\
&+&  S \left[ {\rm Tr} \left(\lambda^* \Delta \bar \Delta +\lambda \Delta^c
\bar \Delta^c
\right) +\lambda'_{ab} {\rm Tr} \left( \Phi_a^T \tau_2 \Phi_b \tau_2
\right)- {\cal M}_R^2 \right] + W'
\end{eqnarray}
where
\begin{equation}
W' = \left[M_\Delta {\rm Tr}(\Delta \bar \Delta) +
M_\Delta^* {\rm Tr}(\Delta^c \bar \Delta^c)\right] + \mu_{ab} {\rm Tr} \left( \Phi_a^T \tau_2 \Phi_b \tau_2 \right) + {\cal M}_S S^2
+ \lambda_S S^3~.
\end{equation}
$Y_{u,d}$ and $Y_{\nu,\ell}$ in Eq. (3) are quark and lepton
Yukawa coupling matrices, while $f$ is the Majorana neutrino
Yukawa coupling matrix.  The $W'$ term listed in Eq. (4) is
optional, in fact when terms in $W'$ are set to zero, the theory
has an enhanced $R$ symmetry.  Under this $R$--symmetry, $\{Q,
~Q^c,~ L,~L^c\}$ fields have charge $+1$, $S$ has charge $+2$, and
all other fields have charge zero with $W$ carrying charge $+2$.
While the general setup of the minimal model includes $W'$, the
special case of $W'=0$ is interesting, as it leads to an
understanding of the $\mu$ term. In the supersymmetric limit, the
VEV of the singlet $S$ is zero, but after SUSY breaking,
$\left\langle S \right\rangle \sim m_{\rm SUSY}$.  Thus the $\mu$
term for the bidoublet $\Phi$ will arise from the coupling
$\lambda'_{ab}$, with a magnitude of order $m_{\rm SUSY}$
\cite{shafi}. It is also in the limit where $W'=0$ that the SUSY
CP problem and the strong CP problem can be explained naturally.
The main difference between the cases $W' \neq 0$ and $W'=0$ from
the low energy perspective is that in the latter case the
left--handed triplet superfields $(\Delta + \overline{\Delta})$
will remain light, also with masses of order $m_{\rm SUSY}$.

The superpotential of Eq. (3)  is invariant under the parity
transformation under which $\Phi\to \Phi^\dagger$,
$\Delta\to\Delta^{c*}$, $\overline{\Delta} \to
\overline{\Delta}^{c*}$, $S \to S^*$, $Q\to Q^{c*}$, $L \to
L^{c*}$, $\theta\to \bar{\theta}$, etc. Parity invariance implies
that the Yukawa coupling matrices $Y_{u,d},~ Y_{\nu,\ell}$ are
hermitian, i.e. $Y_u=Y_u^\dagger$, etc. Additionally,
$\lambda'_{ab}$ are real, as is ${\cal M}_R^2$.  This means that
the effective $\mu$ terms of the bidoublet will be real, provided
that $\left\langle S \right \rangle$ is real. If the $\Phi$ VEVs
are also real, this setup will provide a solution to the SUSY CP
problem and the strong CP problem \cite{rasin,babu}. Below we will
study under what conditions this is achieved and what the
implications of this theory are.

We will work in the ground state corresponding to the following
charge preserving VEV pattern for the triplet
fields.
\begin{eqnarray}\label{VEV}
 \langle\Delta^c \rangle = \left( \begin{array}{cc}
0 & v_R \\
0 & 0
\end{array} \right), \
 \langle\bar \Delta^c\rangle = \left( \begin{array}{cc}
0 & 0 \\
\overline{v}_R & 0
\end{array} \right)~.
\end{eqnarray}
The VEVs of the left--handed triplet fields $(\Delta +
\overline{\Delta})$ are assumed to be zero since no interaction in
the model induces such VEVs. There are two important implications
of this setup: {\bf (i)} Above the parity breaking scale ${\cal
M}_R$, this model has an enhanced global $U(3,c)$ (complexified
$U(3)$) symmetry which is broken by the above VEVs to $U(2,c)$.
This leads to five massless superfields. Three of these
superfields are absorbed by the gauge fields via the super--Higgs
mechanism. There remains two light superfields, which are the
doubly charged Higgs and Higgsino fields $\delta^{c^{--}}$ and
$\overline{\delta}^{c^{++}}$.  These fields will consistently
acquire masses of order TeV or less, as we shall explicitly show
in the next section. Even after soft SUSY breaking terms are
turned on, there is a $U(3)$ symmetry in the potential, which
leads to one massless doubly charged Higgs boson (and its
conjugate).  This field will acquire positive squared mass from
the renormalization group evolution below $v_R$ proportional to
the Bino mass $M_1$. {\bf (ii)} The bi-doublet fields, when
expressed in terms of the components $H_{u,a}, H_{d,a}$ ($a=1,2$),
have a symmetric mass matrix in $W$ due to parity symmetry which
requires $\mu_{12}=\mu_{21}$. (When $W'=0$, $\mu_{ij} =
\lambda'_{ij} \left\langle S \right\rangle$.) Therefore, if we
make one pair of doublets light at the scale $v_R$, it would lead
to vanishing CKM mixing angle. This happens in spite of having two
Yukawa coupling matrices. Consistency then requires that both
pairs of doublets be light below $v_R$.  In this case RGE
extrapolation brings in an asymmetry between $\mu_{12}$ and
$\mu_{21}$. Thus, not only are the potential problems solved by
RGE extrapolation, but the resulting scenario becomes very
predictive.

\vspace*{-0.1in}

\section{Symmetry breaking and the mass of the doubly charged Higgs boson}

To be specific, we will analyze the model with $W'=0$ of Eq. (3).  In the SUSY limit we have
from the vanishing of $D$ and $F$ terms,
\begin{equation}
|v_R| = |\overline{v}_R|,~~\lambda v_R \overline{v}_R = {\cal M}_R^2,~~ \left\langle S \right\rangle = 0~.
\end{equation}
It is easy to determine the VEV of $S$ field that is generated after SUSY breaking.  Only linear
terms in SUSY breaking are relevant for this purpose.  We have
\begin{equation}
V_{\rm soft} = A_\lambda \lambda S {\rm Tr}(\Delta^c
\overline{\Delta}) - C_\lambda {\cal M}_R^2 S + h.c.
\end{equation}
Minimization of the resulting potential yields
\begin{equation}
\left\langle S^* \right \rangle = {1 \over 2|\lambda|} (C_\lambda - A_\lambda)~.
\end{equation}
Note that this is of order $m_{\rm SUSY}$.  If the coupling
$|\lambda|$ is somewhat small, then  $\left\langle S
\right\rangle$ can be above the SUSY breaking scale.  This feature
can be used to make one pair of Higgs doublet superfields somewhat
heavier than the SUSY breaking scale. However, the masses of
doubly charged fermionic fields, which are equal to
$|\lambda|\left\langle S \right \rangle$ must remain below a TeV.
Phenomenology of doubly charged Higgsino has been studied in Ref.
\cite{huitu,han,gunion,other}.

Parity symmetry requires ${\cal M}_R^2$ and $C_\lambda$ be real.
If the trilinear soft breaking terms are proportional to the
corresponding Yukawa coupling matrices,  then we have
$A_{\lambda}$ real as well.  Proportionality will require
$A_\lambda \lambda = A_0 \lambda$, with the universal $A_0$ being
real.   Since the trilinear $A$ terms in the quark sector must be
hermitian by parity, and since the Yukawa coupling matrices are
hermitian, $A_0$ must be real. This condition is realized in many
models of SUSY breaking such as Poloni type supergravity breaking,
gauge mediated SUSY breaking, anomaly mediated SUSY breaking, etc.
We shall adopt this proportionality relation for all the $A$
terms. We see that $\left\langle S \right \rangle$ is then real.
The resulting $\mu$ terms will also be real.  This helps solve the
strong CP problem and the SUSY phase problem.

The full potential of the model relevant for symmetry breaking has $F$
term, $D$ term and soft SUSY breaking contributions.  They are given by
\begin{eqnarray}
V_F &=& \left|\lambda {\rm Tr}(\Delta^c \overline{\Delta}^c +
\lambda'_{ab} {\rm Tr} \left( \Phi_a^T \tau_2 \Phi_b \tau_2
\right)-{\cal M}_R^2\right|^2 + |\lambda|^2 \left|{\rm
Tr}(\Delta^c \Delta^{c \dagger}) + {\rm
Tr}(\overline{\Delta}^c~\overline{\Delta}^{c \dagger}) \right|
\nonumber \\
V_{\rm soft} &=&  M_1^2 {\rm Tr} (\Delta^{c \dagger} \Delta^c) +
M_2^2 {\rm Tr} (\overline{\Delta}^{c \dagger} \overline{\Delta}^c) +M_S^2 |S|^2\nonumber \\
&+& \{A_\lambda \lambda S {\rm Tr}(\Delta^c \Delta^{c \dagger}) -
C_\lambda {\cal M}_R^2 S + h.c.\} \nonumber \\
 V_D &=& {g_R^2 \over
8}\sum_{a}\left|{\rm Tr}(2 \Delta^{c \dagger} \tau_a \Delta^c +
2 \overline{\Delta}^{c \dagger} \tau_a \overline{\Delta}^c + \Phi_a \tau_a^T \Phi_a^\dagger)\right|^2  \nonumber \\
&+& {g'^2 \over 8} \left|{\rm Tr}(2 \Delta^{c \dagger}  \Delta^c +
2 \overline{\Delta}^{c \dagger} \overline{\Delta}^c )\right|^2~.
\end{eqnarray}

Minimizing the potential yields the following two complex
conditions.
\begin{eqnarray}
&~& v_R^*\left[|\lambda|^2|S|^2+M_1^2+g_R^2(|v_R^2
-|\overline{v}_R|^2 + {X \over 2})+g'^2 (|v_R|^2-
|\overline{v}_R|^2)\right]\nonumber \\
&+& \overline{v}_R\left[\lambda A_\lambda S
+ |\lambda|^2(v_R \overline{v}_R-{{\cal M}_R^2
\over \lambda})^*\right] = 0, \nonumber \\
&~& \overline{v}_R^*\left[|\lambda|^2|S|^2+M_2^2-g_R^2(|v_R^2
-|\overline{v}_R|^2 + {X \over 2})-g'^2 (|v_R|^2-
|\overline{v}_R|^2)\right]\nonumber \\
&+& {v}_R\left[\lambda A_\lambda S
+ |\lambda|^2(v_R \overline{v}_R-{{\cal M}_R^2
\over \lambda})^*\right] = 0,
\end{eqnarray}
where we defined $X= \sum_{a=1}^2\left\langle |\phi_1^0|^2 -
|\phi_2^0|^2  \right\rangle_a$. Applying these conditions, we
obtain the following mass squared matrix for the doubly charged
Higgs bosons $(\delta^{c^{-- *}}, \overline{\delta}^{c^{++}})$.
\begin{eqnarray}
{\cal M}^2_{\delta^{++}} = \left(\matrix{-2 g_R^2(|v_R|^2-|\overline{v}_R|^2+{X \over 2})-{\overline{v}_R
\over v_R^*} Y & Y^* \cr Y & 2 g_R^2(|v_R|^2 - |\overline{v}_R|^2+{X \over 2}) - {v_R \over \overline{v}_R^*}Y
}\right)
\end{eqnarray}
where $Y = \lambda A_\lambda S +|\lambda|^2(v_R \overline{v}_R -
{{\cal M}_R^2 \over \lambda})^*$. It is clear that as the $D$ term
is set to zero, there is one massless mode in this sector.
Actually, if $v_R$ is much larger than the SUSY breaking terms,
turning on the $D$ term makes one of the masses negative.  This is
the pseudo--Goldstone boson of the model.  There is no
inconsistency, as this zero  squared-mass will turn positive via
RGE evolution.

Below the scale $v_R$, the mass matrix of the doubly charged Higgs
boson fields has the form
\begin{eqnarray}
{\cal M}^2_{\delta^{++}} = \left(\matrix{M_{++}^2 + \mu_\delta^2 +
\delta_1 & (B\mu)_{\delta} + \delta_{12} \cr (B\mu)_{\delta}^*
+\delta_{12}^* & M^2_{--} + \mu_\delta^2 + \delta_2}\right)
\end{eqnarray}
where $\mu_{\delta} \delta^{++} \delta^{--}$ is the effective
superpotential mass term, $M_{++}^2$ and $M_{--}^2$ are the soft
mass parameters, and $\delta_i$ denote RGE correction factors
corresponding to running from $v_R$ down to the SUSY breaking
scale.  Eq. (12) should match Eq. (11) at $v_R$, which implies
that $M_{++}^2 \simeq M_{--}^2$, $|(B \mu)_\delta| \simeq M_{++}^2
+ \mu_\delta^2$ at $v_R$.  In the large $v_R$ limit, the light
Higgs resulting from Eq. (11) is $(\delta^{*--} -
\delta^{++})/\sqrt{2}$, so the squared mass of this state,
including RGE corrections is $[\delta_1 + \delta_2-2 {\rm
Re}(\delta_{12})]/2$.  There is an upper limit on this mass, which
can be derived as follows.  Let us ignore the off--diagonal entry
for the moment.  The renormalizaion group equation for $M_{++}^2$
has the form \cite{martin}
\begin{equation}
{d M_{++}^2 \over dt} = -{c \over 16 \pi^2} g_1^2 M_1^2 + ...
\end{equation}
where $c = (96/5)$.  Here we have displayed only the positive
contributions to the mass-squared, which would be relevant for
determining the upper limit.  Along with
\begin{equation}
{d g_1 \over dt} = {b_1 \over 16 \pi^2} g_1^3,~~~{dM_1 \over dt} =
{2 b_1 \over 16 \pi^2} g_1^2 M_1^2,
\end{equation}
we can solve for $M_{++}^2$.  In the present model $b_1 = (78/5)$
when the $(\Delta + \overline{\Delta})$ are light, and $b_1 =
12$ when these fields are heavy.  We find
\begin{equation}
M_{++}^2(m_Z) <  {24 \over 5 b_1} M_1^2(m_Z)\left[{\alpha_1^2(v_R)
\over \alpha_1^2(m_Z)}-1\right]~.
\end{equation}
The gauge couplings in this model will remain perturbative up to
about $10^{12}$ GeV when $b_1=(78/5)$ and up to about $10^{14}$ GeV
when $b_1= 12$.  If we choose $\alpha_1(v_R) = 0.1$, we find the
upper limit on $M_{++} < 3.7 M_1$ (for the case where $b_1=12$).  The running of the
$(B\mu)_{\delta}$ will also contribute to the mass of this state,
but this evolution depends on other SUSY breaking parameters.  We
expect the entire contribution to be of order few times $M_1$.

\section{Effective potential and the global minimum of the theory}

Let us now turn attention to the electric charge and/or
$R$--parity breaking global minimum of the model and see how this
problem is  cured by taking loop corrections induced by the heavy
Majorana neutrino Yukawa couplings into account. We will show that
the results of Ref. \cite{kuchi} gets significantly modified,
allowing for the desired charge conserving minimum to be the
global minimum for some domain of the parameters.

It is easy to see why the tree--level potential has a deeper
minimum that violates electric charge and/or $R$--parity.  In the
desired minimum which preserves these quantum numbers, the VEVs of
the triplet fields are as shown in Eq. (5).  Consider the
following alternative VEV configuration.
\begin{eqnarray}
 \langle\Delta^c \rangle = {1 \over \sqrt{2}}\left( \begin{array}{cc}
0 & {v_R} \\
{v_R} & 0
\end{array} \right), \
 \langle\bar \Delta^c\rangle = {1 \over \sqrt{2}} \left( \begin{array}{cc}
0 & {\overline{v}_R} \\
{\overline{v}_R} & 0
\end{array} \right)~.
\end{eqnarray}
This pattern of course breaks electric charge.  All terms in the
scalar potential are exactly the same for this configuration of
VEVs and that of Eq. (5), except in the $SU(2)_R$ $D$--terms.
Since the VEVs  of Eq. (16) are along $\tau_1$, the $D$--terms
vanish for this configuration, while it is nonzero and positive
for the desired configuration.  This proves that the desired VEV
pattern does not correspond to the global minimum of the
potential.

We proceed to compute the Coleman--Weinberg potential of the model by keeping
one family of neutrino Yukawa couplings to the $\Delta^c$ field, as shown by the
$f$ coupling in Eq. (3).  To be able to compare different minima, we use a the
general background with the full $\Delta^c$ and $\overline{\Delta}^c$ fields.
The field--dependent masses of the $(e^c,~\nu^c$) fermionic and scalar fields
can be expressed in terms of the invariant combinations
\begin{eqnarray}
D^2_{1,2} = {1 \over 2} \left[{\rm Tr}(\Delta^{c \dagger} \Delta^c) \pm \sqrt{ {\rm Tr}(\Delta^{c \dagger} \Delta^c)^2
-{\rm Tr}(\Delta^c \Delta^c) {\rm Tr} (\Delta^{c \dagger} \Delta^{c \dagger})}\right]~.
\end{eqnarray}
We also define $\overline{D}_{1,2}^2$ in an analogous way, with the replacement of $\Delta^c$ by
$\overline{\Delta}^c$ in Eq. (17).
Including the soft SUSY breaking contributions, the $F$--term contributions, and the $D$--term
contributions, the field--dependent masses of the sleptons $(\tilde{e^c},~\tilde{\nu^c})$, and
the corresponding fermions are
found to be
\begin{eqnarray}
m_{1,2}^2 &=& |f|^2 D_1^2 + m_{L^c}^2 + {g_R^2 \over 2}[(D_2^2 -
\overline{D}_2^2) - (D_1^2 - \overline{D}_1^2)] -{{g'^2} \over
2}[(D_1^2 - \overline{D}_1^2) +
(D_2^2 - \overline{D}_2^2)], \nonumber \\
&\pm& \left|A_f f D_1 + \lambda^* S^* f \overline{D}_1 \right|^2
\nonumber \\
m_{3,4}^2 &=& |f|^2 D_2^2 + m_{L^c}^2 + {g_R^2 \over 2}[(D_1^2 -
\overline{D}_1^2) - (D_2^2 - \overline{D}_2^2)] -{{g'^2} \over
2}[(D_1^2 - \overline{D}_1^2) +
(D_2^2 - \overline{D}_2^2)], \nonumber \\
&\pm& \left|A_f f D_2 + \lambda^* S^* f \overline{D}_2 \right|^2 \nonumber \\
m_{F_1}^2 &=& |f D_{1}|^2,~\nonumber\\
m_{F_2}^2 &=& |f D_{2}|^2.
\end{eqnarray}
Here the $m_{1-4}$ correspond to the masses of the four real
scalar states, while $m_{F_{1,2}}$ are the masses of the two
fermionic states.

With these mass eigenvalues, one can compute the effective potential in the Landau gauge
in the $\overline{DR}$ scheme from the expression
\begin{equation}
V_{\rm eff}^{\rm 1-loop} = {1 \over 64 \pi^2}\sum_i(-1)^{2s}
(2s+1) M_i^4 \left[{\rm Log}({M^2_i \over \mu^2}) - {3 \over
2}\right] ~.
\end{equation}
We expand this potential in the limit where SUSY breaking parameters are small compared to
the VEVs of the $(\Delta^c, \overline{\Delta}^c)$ fields.  In the SUSY limit, vanishing of the
$D$--terms require $D_1^2 = \overline{D}_1^2,~D_2^2 = \overline{D}_2^2$.  So we use the expansion
\begin{equation}
\overline{D}_1^2 - D_1^2 = a_1 m_{L^c}^2,~~~~~\overline{D}_2^2 -
D_2^2 = a_2 m_{L^c}^2
\end{equation}
where $m_{L^c}^2$ denotes the soft SUSY breaking mass of the
slepton doublet.  Defining
\begin{equation}
x = {{{\rm Tr}(\Delta^c \Delta^c){\rm Tr}(\Delta}^{c \dagger}
\Delta^{c \dagger}) \over [{\rm Tr}(\Delta^{c \dagger}
\Delta^c)]^2}
\end{equation}
we find the leading contribution to $V_{\rm eff}$ to be {\small
\begin{eqnarray}
\hspace*{-0.2in} V_{\rm eff}^{\rm 1-loop} &=&
 -{|f|^2 m_{L^c}^2 {\rm
Tr}(\Delta^c {\Delta}^{c \dagger}) \over 64 \pi^2} \left
[(4+2~{\rm ln}2)+2 (a_1-a_2)g_R^2\sqrt{1-x}+ 2(a_1+a_2)g'^2+
\right.
\nonumber \\
&-& \left\{2+(a_2-a_1) g_R^2 +
(a_2+a_1)g'^2\right\}\left(1-\sqrt{1-x}\right){\rm ln}
\left({|f|^2 {\rm Tr}(\Delta^c {\Delta}^{c \dagger}) \over 2
\mu^2}\left(1-\sqrt{1-x}\right)\right) \nonumber \\
&+& \left\{\left((a_2-a_1) g_R^2 -
(a_2+a_1)g'^2\right)\left(1+\sqrt{1-x}\right)-2\sqrt{1-x}\right\}{\rm
ln} \left({|f|^2 {\rm Tr}(\Delta^c {\Delta}^{c \dagger})
\over 2 \mu^2}\left(1+\sqrt{1-x}\right)\right) \nonumber \\
&-&\left.  2~{\rm ln}\left({|f|^2 {\rm Tr}(\Delta^c {\Delta}^{c
\dagger}) \over \mu^2}\left(1+\sqrt{1-x}\right)\right) \right]
\end{eqnarray}}

\begin{figure}[h]
\begin{center}
\begin{tabular}{cc}
\includegraphics[width=0.5\linewidth]{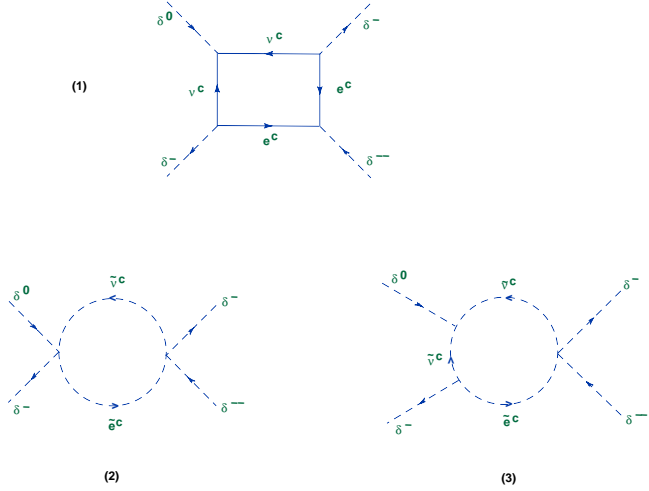}
\end{tabular}
\vspace*{0.1in} \caption{Diagrams inducing effective quartic
coupling of Eq. (21).}
\end{center}
\end{figure}

Clearly, these loop contributions vanish in the SUSY limit.  The
non--vanishing terms arise because the cancelation between the
first two diagrams of Fig. 1 is no longer exact, once SUSY
breaking is turned on. And diagram (c) has no fermionic
counterpart. The most interesting aspect of the one--loop
effective potential is the appearance of the structure ${\rm
Tr}(\Delta^c \Delta^c){\rm Tr}({\Delta}^{c \dagger} {\Delta}^{c
\dagger})$, which was absent in the tree--level potential.  If we
make a further expansion in small $x$, Eq. (22) will result in the
following quartic coupling:
\begin{eqnarray}
V^{\rm quartic} &=&  -{|f|^2 m_{L^c}^2 {\rm Tr}(\Delta^c
\Delta^c){\rm Tr}({\Delta}^{c \dagger} {\Delta}^{c \dagger}) \over
128 \pi^2|v_R|^2} \left [\{2-\{a_1-a_2)g_R^2-(a_1+a_2)g'^2\}(1+
2~{\rm ln}2) \right.
\nonumber \\
&+& \left. (a_1-a_2)g_R^2 ~{\rm ln}{|fv_R|^2 \over \mu^2} -
\{2-(a_1-a_2)g_R^2+(a_1+a_2)g'^2\}~{\rm ln}x\right] + ...
\end{eqnarray}
\noindent where the ... indicates higher order terms in $x$ and
$x$--independent terms.  In the desired vacuum we have
$D_2=\overline{D}_2 =0$, so that the coefficient $a_2$ is zero.
Minimization conditions (Eq. (10)) determine $a_1$ as
\begin{equation}
(g_R^2 + g'^2) a_1 m_{L^c}^2 \simeq {1 \over 2} (M_1^2 - M_2^2 +g_R^2
X)~,
\end{equation}
where $M_{1,2}^2$ are the soft mass squared of the $(\Delta^c,~
\overline{\Delta}^c$) fields.  In supergravity type SUSY breaking,
one would expect $M_1^2 \leq M_2^2$, as $\Delta^c$  has the
Majorana Yukawa coupling which would lower its mass from the
universal mass, while $\overline{\Delta}^c$ does not. Using this
we find that for the charge conserving vacuum to be lower than the
charge breaking vacuum, we would need $m_{L^c}^2$ to be negative.
In such a situation, we can derive upper limits on the stau
masses.  (We assume that the third family fermions have the
largest Majorana Yukawa coupling $f$.)  Note that the positive
contributions to the masses of $\tilde{\tau}_R$ and
$\tilde{\tau}_L$ arise from the gaugino masses $M_1$ and $M_2$ \cite{martin}.
\begin{eqnarray}
16 \pi^2 {d m^2_{\tilde{\tau}_R}\over dt} &=& -{24 \over 5} g_1^2
M_1^2 + ...\nonumber
\\
16 \pi^2 {d m^2_{\tilde{\tau}_L}\over dt} &=& -{6 \over 5} g_1^2
M_1^2 - 6 g_2^2 M_2^2 + ...
\end{eqnarray}
where the ... denote terms that would decrease the scalar mass in
the evolution from $v_R$ to $m_Z$.  We have $b_2= 6$ when $(\Delta
+ \overline{\Delta})$ fields are light, and $b_2=2$ when they are
heavy.  The upper limits on the stau masses are found to be
\begin{eqnarray}
M_{\tilde{\tau_R}}^2(m_Z) &<&  {6 \over 5 b_1}
M_1^2(m_Z)\left[{\alpha_1^2(v_R) \over
\alpha_1^2(m_Z)}-1\right]~,\nonumber \\
M_{\tilde{\tau_L}}^2(m_Z) &<&  {3 \over 10 b_1}
M_1^2(m_Z)\left[{\alpha_1^2(v_R) \over \alpha_1^2(m_Z)}-1\right]
+{3 \over 2 b_2} M_2^2(m_Z)\left[{\alpha_2^2(v_R) \over
\alpha_2^2(m_Z)}-1\right] .
\end{eqnarray}
Both these limits are in th acceptable range. For $\alpha_1(v_R) =
0.1$, we find the right--handed stau mass to be bounded by about
1.9 $M_1$ (for $b_1=12$), with the left--handed stau roughly two times heavier.

\section{CKM angles out of radiative corrections}

As noted earlier, our model predicts that the
CKM angles vanish at the tree--level due to left--right symmetry. The reason for
this is that the $2\times 2$ ($H_u,~H_d$) Higgsino mass matrix is
symmetric.  When one pair of light MSSM Higgs superfields is extracted from
such a symmetric matrix, it follows that the up and down quark Yukawa coupling
matrices to these light doublets will be the same.  This is assuming that only
one pair of doublets survives below $v_R$. Therefore once
electroweak symmetry breaks, we have $M_u~=~\xi M_d$ and hence
$V_{CKM}~=~1$. Consistency with CKM mixings then requires that both pairs of
Higgs doublets remain light below $v_R$.  In that case, below
$v_R$, the bidoiblet mass terms $\mu_{ab}$
will receive asymmetric radiative RGE corrections, in the momentum range $v_R$ to $\mu_\Phi$,
because parity is violated in this regime. (We denote the scale of the heavy doublet
mass as $\mu_\Phi$.) To leading
order the quark Yukawa couplings do not induce an asymmetry in $\mu_{ab}$.
However, since the right--handed neutrinos decouple
below  $v_R$, the lepton sector induces an asymmetry.
Only the charged lepton Yukawa couplings contribute to the evolution of $\mu_{ab}$, making the RGE
contribution to $\mu_{12}$ different from that of $\mu_{21}$. As a
result, when the $H_u,H_d$ mass matrix is diagonalized at a scale $\mu_\Phi$ below $v_R$,
so that only one pair of Higgs superfields remain
light, the resulting light Higgs doublets couple to up and down quarks with
different Yukawa coupling matrices.

The RGE for the asymmetry between $\mu_{12}$ and  $\mu_{21}$
(to leading order) is
\begin{eqnarray}
{d  \over dt}(\mu_{12} - \mu_{21}) = {\mu_{12} + \mu_{21} \over 32
\pi^2}{\rm Tr}(Y_\nu^\dagger Y_\nu - Y_\ell^\dagger Y_\ell),
\end{eqnarray}
which can be solved to determine the asymmetry in $\mu_{ij}$.  We
obtain $(\mu_{12}-\mu_{21})/(\mu_{12}+\mu_{21}) \simeq 1/(16 \pi^2){\rm
Tr}(Y_\nu^\dagger Y_\nu - Y_\ell^\dagger Y_\ell){\rm
ln}(v_R/\mu_\Phi)$, where $\mu_\Phi$ is the mass of the heavy bidoublet. The
suppression factor that apperas in the CKM angles is about 0.1
when one of the leptonic Yukawa coupling entries is
of order one.  This can lead to reasonable values for the CKM
angles.

\section{FCNC, the strong CP and the SUSY CP problems}

The presence of a second pair of Higgs doublets coupling to
fermions implies that there will be tree--level flavor changing
neutral currents mediated by the Higgs.  Experimental constraints
will require that one pair of Higgs doublets be heavy, with mass
of the order of few to 50 TeV \cite{ji,pospelov}.  This can be seen from the mass matrices
of the quarks,
\begin{eqnarray}
M_u &=& Y_u \kappa_u + Y_d \kappa_u' \nonumber \\
M_d &=& Y_u \kappa_d' + Y_d \kappa_d
\end{eqnarray}
where $\kappa_i$ are the VEVs of the neutral components.  These
equations can be used to solve for the Yukawa coupling matrices.
For example, $Y_d = (\kappa_u M_u - \kappa_d' M_d)/(\kappa_u
\kappa_d - \kappa_u' \kappa_d')$.  In a basis where $M_d$ is
diagonal, $M_u = \hat{V}^T D_u \hat{V}^*$, where $\hat{V} =
P.V.Q$, with $V$ being the CKM matrix in the standard
parametrization, and $P,Q$ being phase matrices.  $D_u$ is the
diagonal up--quark mass matrix.  Flavor changing Higgs couplings
can be then readily derived:
\begin{equation}
{\cal L}^{FCNC} = \left({\kappa_u \over \kappa_u \kappa_d -
\kappa_u' \kappa_d'}\right)Q_i Q_j^* (D_u)_kV_{ki} V_{kj}^* H^0 +
h.c.
\end{equation}
Due to the hermiticity of this matrix, the unknown phase matrix
$Q$ disappears from processes such as $\epsilon_K$.  We find
stringent limit on the mass of $H_0$, $m_{H^0} \geq (30-50)$ TeV, if
there is no cancelation between the Higgs exchange and the SUSY
squark--gluino exchange box diagram.  If such cancelations are allowed,
the limit on $H^0$ mass is considerably reduced \cite{ji}.  As noted
after Eq. (8), the model allows for one pair of Higgs doublets to be naturally
heavier than the SUSY breaking scale, thus satisfying the FCNC constraint.

Since two pairs of Higgs doublets must survive below $v_R$, there
are calculable FCNC via SUSY diagrams.  The most significant ones
are the gluino box diagram for $K^0 - \overline{K}^0$ mixing.  We
find that these constraints are met in the model.

The basic idea behind parity as a solution for the strong CP
problems is that left--right symmetry leads to hermitian Yukawa
couplings \cite{senj}.  If the VEVs of bi--doublet Higgs fields are
real, this would
lead to a solution to the strong CP problem. The reality of the
VEVs is not guaranteed by parity and always involves additional
assumptions. Supersymmetry provides this extra symmetry in minimal
left--right models without any singlet fields as shown in
\cite{rasin1}. When there are gauge singlet fields in the theory,
this needs to be reinvestigated.  As we noted in the symmetry
breaking discussion, if $W'=0$, which can be enforced by an $R$--symmetry,
one can have a scenario where the singlet VEV is real.
In such a setup not only is the strong CP problem solved, but the
weak SUSY CP problem is also solved.  The EDMs of the electron and
the neutron will be vanishing due to parity at the scale $v_R$.
Renormalization group evolution does induce small EDMs, but well
within experimental limits.

\section{Conclusion}

In conclusion, we have pointed out that the minimal renormalizable
supersymmetric left-right model is completely consistent
phenomenologically  without any need for higher dimensional
operators or spontaneous $R$--parity violation. The scale of
left--right symmetry can now be higher than TeV. The model can
solve the strong CP problem without fear of large contributions to
$\overline{\theta}$ from non-renormalizable terms (since they are now not
needed). The model also provides a simple solution based on parity
symmetry for the SUSY CP problem.  The effective potential of the theory, which has
important contributions from heavy Majorana Yukawa couplings, allows for the
charge conserving and $R$--parity conserving minimum to be the
global minimum.  The model predicts light (sub--TeV) doubly
charged Higgs bosons and their superpartners.

\vspace*{-0.1in}

\section*{Acknowldgements}
We wish to thank Zurab Tavartkiladze for discussions.  KSB is supported in part by
DOE grants DE-FG02-04ER41306 and DE-FG02-ER46140. RNM is supported by NSF grant No. PHY-0652363.



\end{document}